\documentclass[11pt,twoside]{article}
\usepackage{asp2004}
\usepackage{psfig}
\usepackage{epsf}
\usepackage{graphics}
\usepackage{lscape}
\def\edcomment#1{\iffalse\marginpar{\raggedright\sl#1\/}\else\relax\fi}
\reversemarginpar

\begin{document}
\title{Direct Constraints
on the Properties and Evolution of Dark Energy}
\author{Ruth A. Daly}
\affil{Department of Physics, 
Berks-Lehigh Valley College, \\Penn State University, Reading, PA 19610}
\author{S. G. Djorgovski}
\affil{Division of Physics, Mathematics, and Astronomy, \\California Institute of Technology, MS 105-24, Pasadena, CA 91125}

\begin{abstract}

We describe a method to derive the expansion and acceleration rates
directly from the data, without the need for the specification of a
theory of gravity, and without adopting an a priori parameterization
of the form or redshift evolution of the dark energy.
If one also specifies a theory of gravity we can also determine the
pressure, energy density, and equation of state of the dark energy as
functions of redshift.
We then apply this methodology on a modern data set of distances
to Supernovae and to radio galaxies.
We find that the universe transitions from deceleration to acceleration
at a redshift of $z_T \approx 0.4$, and the present value of deceleration
parameter is $q_0 = -0.35 \pm 0.15$.  The standard ``concordance model''
with $\Omega_0 = 0.3$ and $\Lambda = 0.7$
provides a reasonably good fit to the dimensionless expansion rate as a
function of redshift, though it fits the dimensionless acceleration rate
as a function of redshift less well.
Adopting General Relativity as the theory of gravity, we obtain the
redshift trends for the pressure, energy density, and equation of state
of the dark energy out to $z \sim 1$.  They are generally consistent with
the concordance model, at least out to $z \sim 0.5$, but the existing
data preclude any stronger conclusions at this point.  For the present
values of these quantities we obtain
$p_0 = -0.6 \pm 0.15$,
$f_0 = 0.62 \pm 0.05$, and
$w_0 = -0.9 \pm 0.1$.
Application of this methodology to richer data sets in the future may
provide valuable new insights into the physical nature and evolotion
of the dark energy.
\end{abstract}
\thispagestyle{plain}

\section{Introduction}
One way to determine the expansion and acceleration rates of the
universe as functions of redshift is through studies of the coordinate
distance to sources at different redshift.  This has been 
accomplished with 
a variety of techniques including the use of type Ia supernovae 
(e.g. Riess et al. 1998; Perlmutter et al. 1999; Tonry et al. 2003; 
Knop et al. 2003; Barris et al. 2004; Riess et al. 2004) 
and powerful classical double 
radio galaxies (e.g. Daly 1994; Guerra \& Daly 1998; Guerra, Daly, \& Wan 2000;
Podariu et al. 2003).  

Luminosity or coordinate distances can then be used 
to study the
expansion history of the universe and the properties of the dark
energy in a variety of ways.  The analysis techniques
fall into
two broad categories: the integral approach and the differential approach.
The former, traditional approach involves the integration
of a theoretically predicted expansion rate
over redshift to obtain predicted coordinate distances to different
redshifts; the difference between these predicted coordinate distances
and the observed coordinate distances is then minimized to obtain
the best fit model parameters.  This approach usually
requires the specification
of a theory of gravity (generally taken to be General Relativity; GR) and
a parameterization of the redshift evolution of the dark energy.
Maor, Brustein, \& Steinhardt (2001) and Barger \& Marfatia (2001)
discuss the difficulties involved with the use of this method
to extract the properties and redshift behavior of the dark
energy.  Different approaches have
been developed to extract the redshift behavior of the dark
energy using the integral method (e.g. Starobinsky 1998; Huterer
\& Turner 1999, 2001; Saini et al. 2000; Chiba \& Nakamura 2000;
Maor, Brustein, \& Steinhardt 2001; Golaith et al. 2001;
Wang \& Garnavich 2001;
Astier 2001; Gerke \& Efstathiou 2002; Weller \& Albrecht 2002;
Padmanabhan \& Choudhury 2002; Tegmark 2002; Daly \& Guerra 2002; 
Huterer \& Starkman 2003;  Sahni et al. 2003; Alam et al. 2003;
Podariu et al. 2003; 
Wang \& Freese 2004; Wang et al. 2004; Wang \& Tegmark; 
Nessier \& Perivolaropoulos 2004;
Gong 2004a,b; Zhu, Fujimoto, \& He 2004; Elgaroy \& Multamaki 2004;
Huterer \& Cooray 2004; Alam, Sahni, \& Starobinsky 2004).

\section{The Methodology}

The differential approach has been investigated by
Daly \& Djorgovski (2003, 2004a), and is briefly summarized
here.  It is well known (e.g. Weinberg 1972; Peebles 1993;
Peebles \& Ratra 2003)
that the dimensionless expansion rate $E(z)$
can be written as the derivative of the dimensionless coordinate
distances $y(z)$; the dimensionless coordinate distance $y(z)$ is
simply related to the coordinate distance $(a_0r)$ through the
equation $y = (H_0/c)(a_0r)$, where $c$ is the speed of
light and $H_0$ is Hubble's constant.  The
expression for $E(z)$ is particularly simple when
the space curvature term is equal to zero.  In this case,
\begin{equation}
\left( { \dot{a} \over a} \right)~H_0^{-1} \equiv
E(z) = (dy/dz)^{-1}~,
\label{eofz}
\end{equation}
where $a$ is the cosmic scale factor.  
This representation follows directly
from the Friedman-Robertson-Walker line element, and does not
require the specification of a theory of gravity.  Similarly,
in a spatially flat universe 
(which is supported by CMBR measurements,
Spergel et al. 2003),
it is shown
in Daly \& Djorgovski (2003) that the dimensionless deceleration parameter
\begin{equation}
- \left({\ddot{a} a \over \dot{a}^2}\right)
\equiv q(z) = - [1+(1+z)(dy/dz)^{-1}~d^2y/dz^2]
\label{qofz}
\end{equation}
also follows
directly from the FRW line element, and is independent of
any assumptions regarding the dark energy or a
theory of gravity.  Thus, measurements of the dimensionless
coordinate distance to sources at different redshifts can be used
to determine $dy/dz$ and $d^2y/dz^2$, which can then be used to
determine $E(z)$ and $q(z)$, and these direct measures 
are completely model-independent, as discussed by Daly \&
Djorgovski (2003).

In addition, if a theory of gravity is specified, the measurements of
$dy/dz$ and $d^2y/dz^2$ can be used to determine the pressure, energy
density, and equation of state of the dark energy as functions of
redshift (Daly \& Djorgovski 2004a,b); we assume the standard GR for
this study.
{\it These determinations are completely independent of
any assumptions regarding the form or properties of the dark
energy or its redshift evolution.}
Thus, we can use the data to determine these functions directly,
which provides an approach that is complementary to the standard
one of assuming a physical model, and then fitting the
parameters of the chosen function.

In a spatially flat, homogeneous, isotropic universe with non-relativistic
matter and dark energy
Einstein's equations are
$({ \ddot{a} / a} ) = -{(4 \pi G / 3})~
(\rho_m + \rho_{DE} + 3 P_{DE})$
and
$({ \dot{a} / a} )^2 = ({8 \pi G / 3})~ (\rho_m + \rho_{DE})~,$
where $\rho_m$ is the
mean mass-energy density of non-relativistic matter,
$\rho_{DE}$ is the mean mass-energy
density of the dark energy, and $P_{DE}$ is the pressure of the dark energy.
Combining these equations, we find
$(\ddot{a}/a)=-0.5[(\dot{a}/a)^2~+(8 \pi G)~P_{DE}]$.

Defining the critical density at the present
epoch in the usual way,
$\rho_{0c} = 3H^2_0/(8 \pi G)$, it is easy to show that
$p(z) \equiv ({P_{DE}(z) / \rho_{0c}}) =
({E^2(z) / 3})~[2q(z)-1]~.$
Combining this expression with eqs. (1) and (2) we
obtain the pressure of the dark energy as a function of redshift
in terms of first and second derivatives of the dimensionless
coordinate distance $y$ (Daly \& Djorgovski 2004a)
\begin{equation}
p(z) = -(dy/dz)^{-2}[1+(2/3)~(1+z)~(dy/dz)^{-1}~(d^2y/dz^2)]~.
\label{pofz}
\end{equation}
Thus, the pressure of the dark energy can be determined
directly from measurements of the coordinate distance.
In addition, this provides a
direct measure of the cosmological constant for
Friedmann-Lemaitre models since in these
models $p = -\Omega_{\Lambda}$.  If more than one new component is
present, this pressure is the sum of the pressures of the new
components.

Similarly, the energy density of the dark energy can be obtained
directly from the data
\begin{equation}
f(z) \equiv \left( {\rho_{DE}(z) \over \rho_{0c}} \right)
= (dy/dz)^{-2} - \Omega_{0}(1+z)^3~,
\label{fofz}
\end{equation}
where $\Omega_{0} = \rho_{0m}/\rho_{0c}$ is the fractional contribution
of non-relativistic matter to the total critical density at zero redshift,
and it is assumed that this non-relativistic matter evolves as $(1+z)^3$.
If more than one new component is present, then $f$ includes the
sum of the mean mass-energy densities of the new components.

The equation of state $w(z)$ is defined to be the ratio of the pressure of
the dark energy to it's energy-density $w(z) \equiv P_{DE}(z)/\rho_{DE}(z)$.
As shown by Daly \& Djorgovski (2004a), the equation of state is
\begin{equation}
w(z) = -{[1 + (2/3)
~(1+z)~(dy/dz)^{-1}~(d^2y/dz^2)] \over [1-(dy/dz)^2~\Omega_{0}~(1+z)^3]}~.
\label{wofz}
\end{equation}
Here, $w$ is the equation of state of the dark energy; if more than one
new component contributes to the dark energy, $w$ is the ratio of the sum
of the total pressures of the new components to their total mean mass-energy
densities.

\begin{figure}[!ht]
\plotfiddle{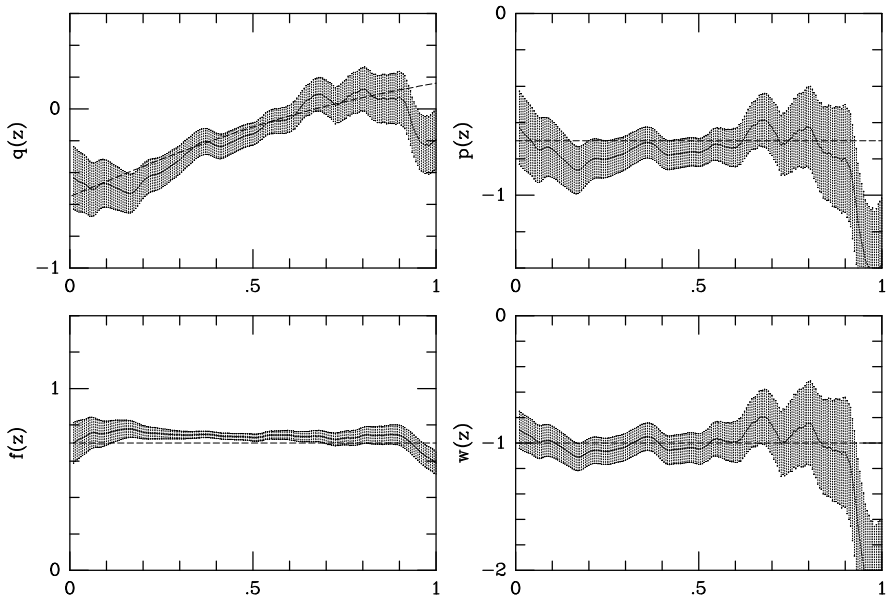}{3.0in}{-90}{50}{50}{-200}{250}
\caption{Application of our methods 
on the simulated (pseudo-SNAP)
data set, obtained with equations (2), (3), (4), and (5) 
respectively   
as described in the text, using a window function with
$\Delta z = 0.4$.  The dotted/hatched regions show the recovered trends
for the quantities of interest.  The assumed cosmology is a standard
Friedmann-Lemaitre model with $\Omega_0 = 0.3$ and $\Lambda_0 = 0.7$,
and the theoretical (noiseless) values of the measured quantities are
shown as dashed lines.  There is a good correspondence (typically well
within $\pm 1 \sigma$) up to $z \sim 0.9$, except in the case of $f(z)$
where a small systematic bias is present, and the formally evaluated
errors may be too small as an artifact of the numerical procedure.
}
\end{figure}

\begin{figure}[!ht]
\vspace{0in}
\plotone{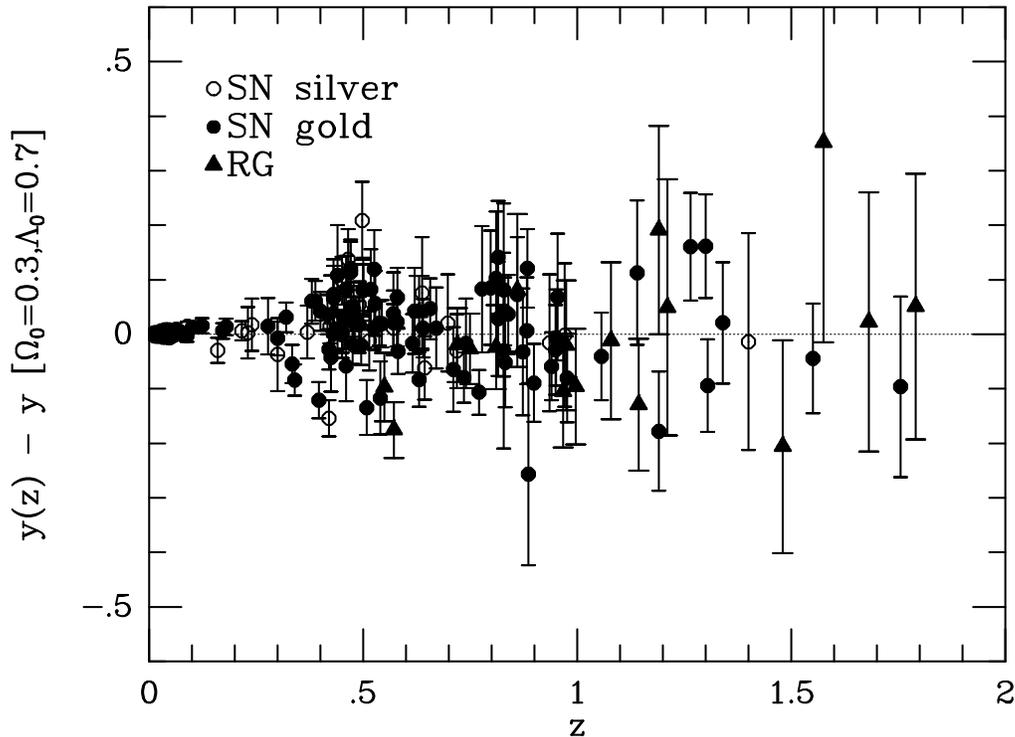}
\caption{The difference between the dimensionless coordinate
distances and those expected in a spatially flat universe
with a cosmological constant and $\Omega_{0} = 0.3$.
SNe and RGs are plotted with different symbols as indicated.
There is no significant systematic offset between them in
the redshift range where there is an overlap.
}
\end{figure}

\begin{figure}[!ht]
\vspace{0in}
\plotone{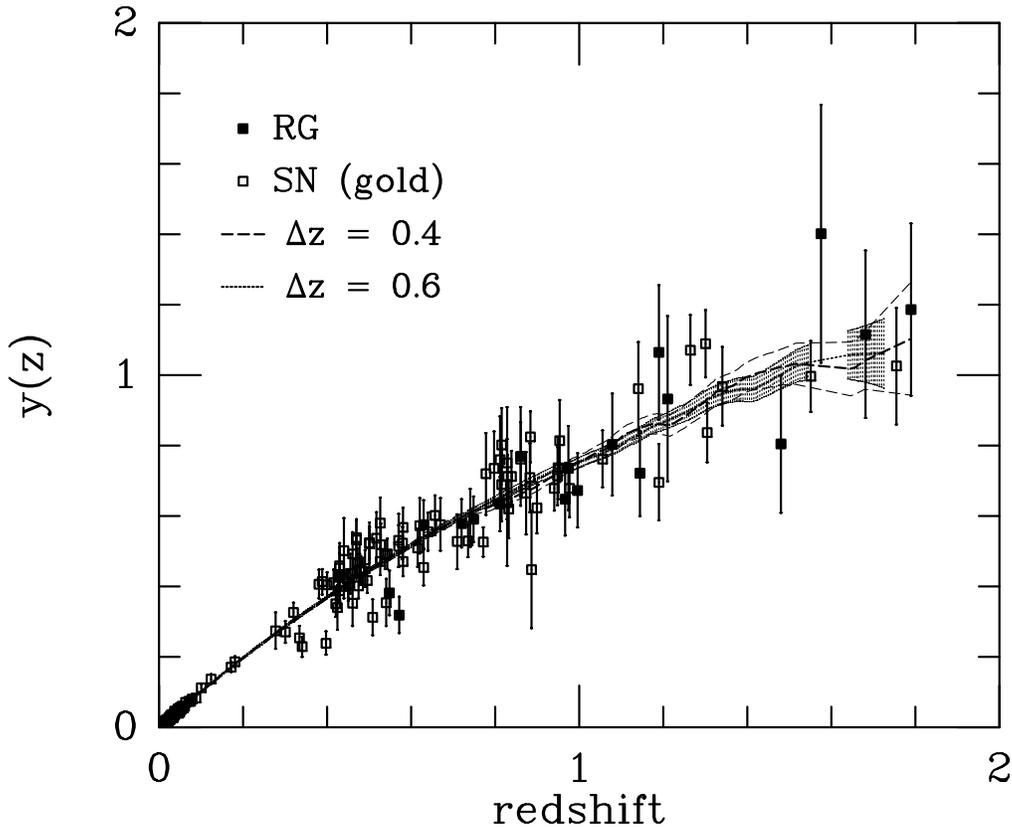}
\caption{Dimensionless coordinate distances $y(z)$ to 20 radio
galaxies and the  ``gold'' sample SNe as a function
of  z. The smoothed values of y along with their
1 $\sigma$ error bars obtained for window function
widths $\Delta z = 0.4$  (dashed lines)
and 0.6 (dotted line and hatched error range) are also shown.
Note again that the new high-redshift SNe values agree quite well
with those of the high-redshift RGs.
}
\end{figure}

\begin{figure}[!ht]
\vspace{0in}
\plotone{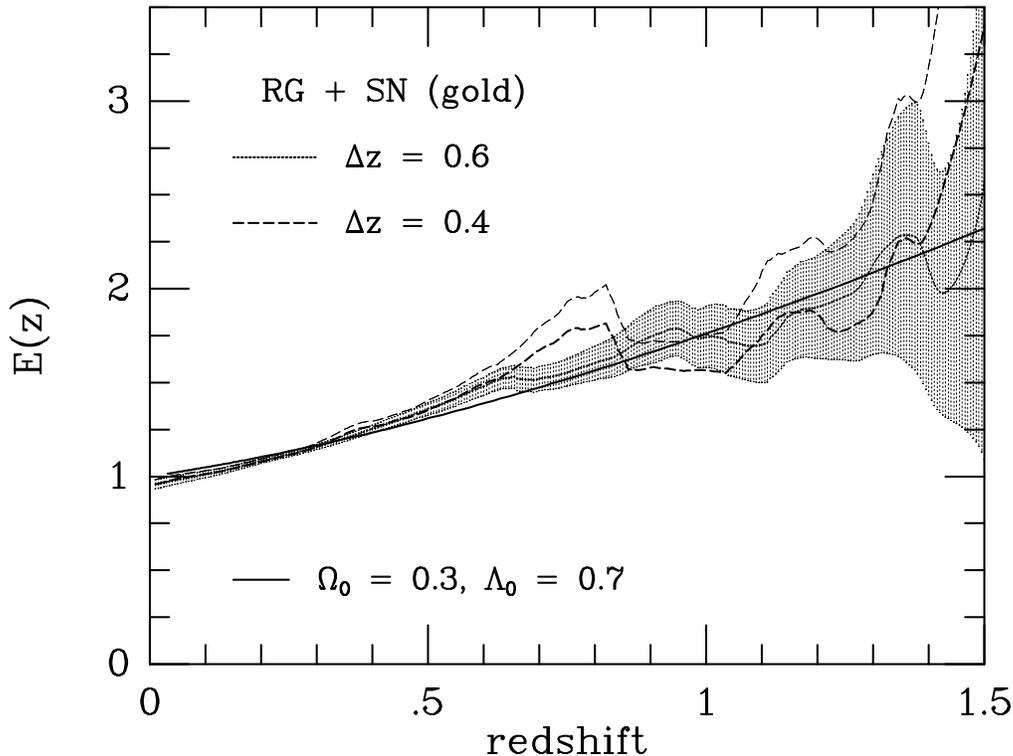}
\caption{The derived values of the dimensionless expansion rate
$E(z) \equiv (\dot{a}/a)H_0^{-1}=(dy/dz)^{-1}$
obtained with window functions of width $\Delta z = 0.4$
and their 1 $\sigma$ error bars
(dashed lines) and 0.6 (dotted line and hatched error range).
At a redshift of zero, the value of $E$ is $E_0 = 0.97 \pm 0.03$.
The value of $E(z)$ predicted in a spatially flat universe with
a cosmological constant and $\Omega_0 =0.3$ is also shown, and
provides a reasonable fit to the data.}

\end{figure}

\begin{figure}[!ht]
\vspace{0in}
\plotone{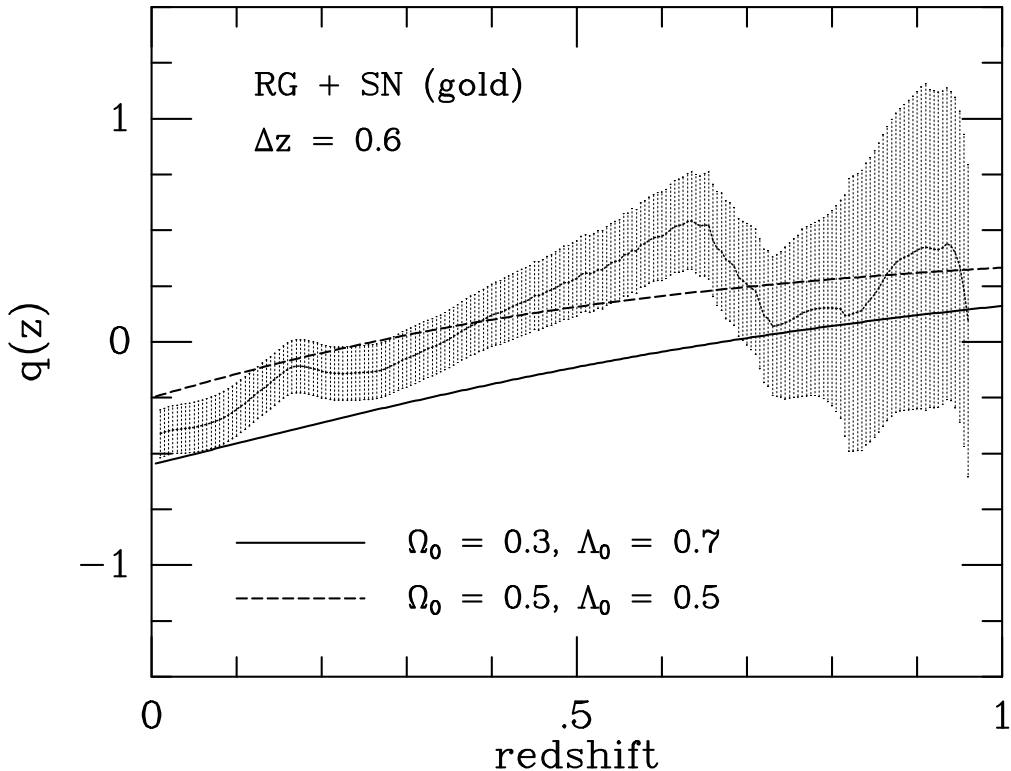}
\caption{The derived values of deceleration parameter $q(z)$
(see equation 2) 
and their 1 $\sigma$ error bars
obtained with window function of width $\Delta z = 0.6$
applied to the RG plus gold SNe
sample.  The universe transitions from acceleration
to deceleration at a redshift $z_T \approx 0.4$.
The value of the deceleration parameter at zero redshift
is $q_0 = -0.35 \pm 0.15$.  Note that this
determination of $q(z)$ only depends upon the assumptions that the
universe is homogenous, isotropic, expanding, and spatially flat,
and it does not depend on any assumptions about the nature of
the dark energy, or the correct theory of gravity.
Solid and dashed lines show the expected dependence in the standard
Friedmann-Lemaitre models with zero curvature, for two pairs of
values of $\Omega_0$ and $\Lambda_0$.
}

\end{figure}

\begin{figure}[!ht]
\vspace{0in}
\plotone{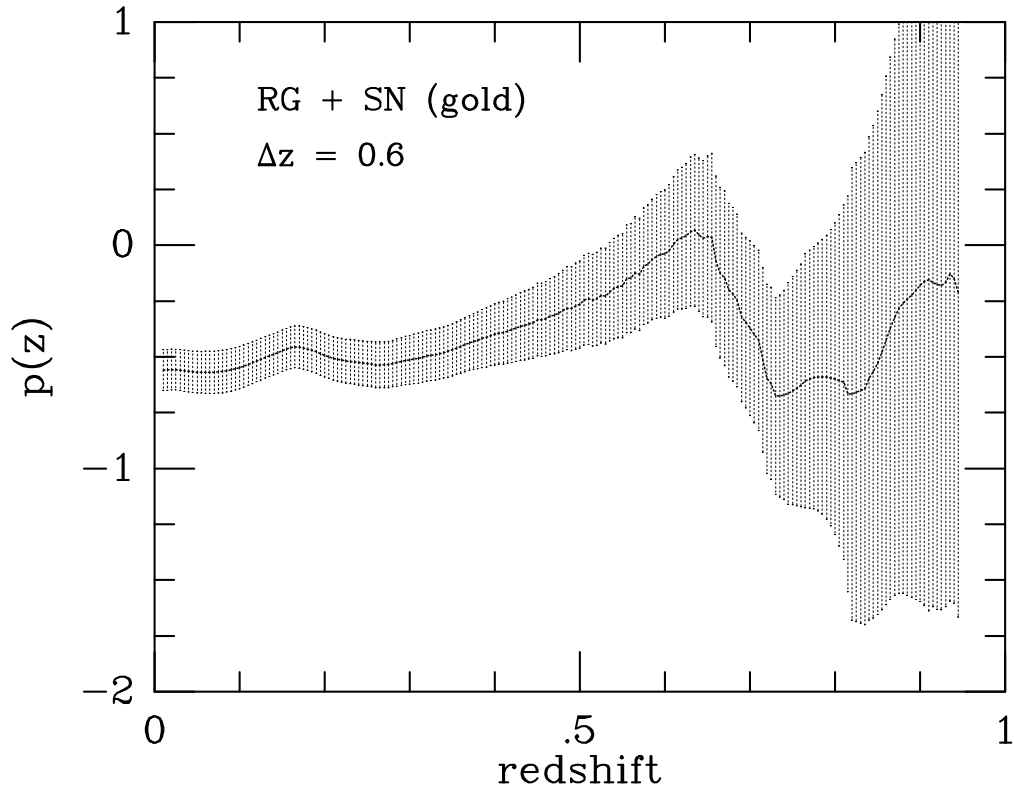}
\caption{The derived values of dark energy pressure $p(z)$
(see equation 3),
obtained with window function of width $\Delta z = 0.6$.
This derivation of $p(z)$ requires a choice of theory of gravity,
and General Relativity has been adopted here.
The value at zero redshift is $p_0 = -0.6 \pm 0.15$.
}
\end{figure}

\begin{figure}[!ht]
\vspace{0in}
\plotone{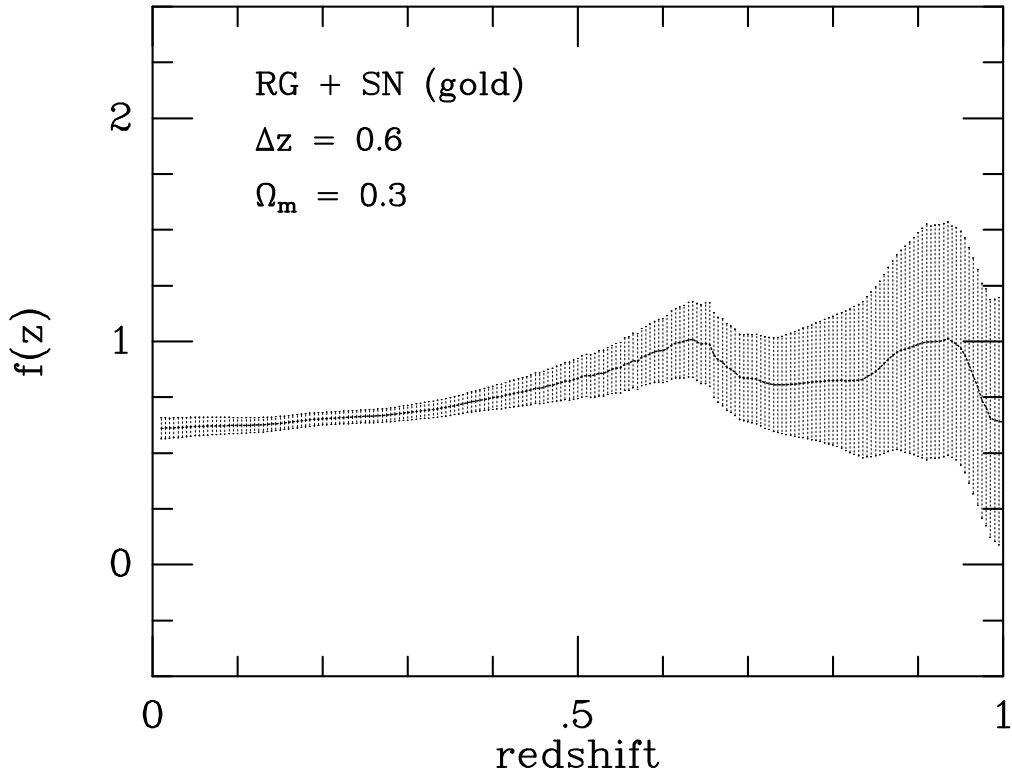}
\caption{The derived values of the dark energy density fraction
$f(z)$ (see equation 4), 
obtained with window function of width $\Delta z = 0.6$.
This derivation of $f(z)$ requires of theory of gravity
and the value of $\Omega_{0}$ for the nonrelativistic matter;
General Relativity has been adopted here, and
$\Omega_0 = 0.3$ is assumed. The value at zero
redshift is $0.62 \pm 0.05$.
}

\end{figure}

\begin{figure}[!ht]
\vspace{0in}
\plotone{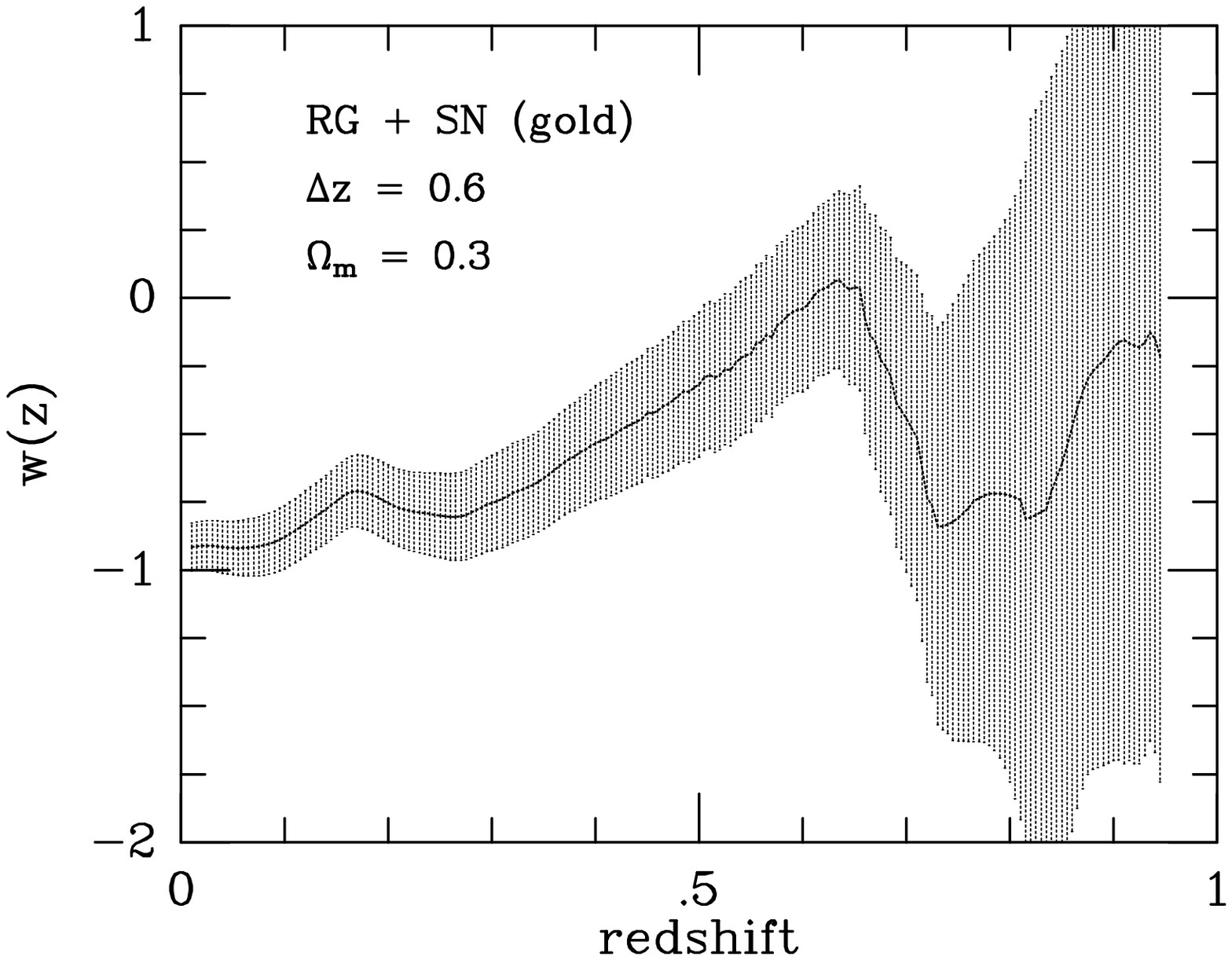}
\caption{The derived values of the dark energy equation of state
parameter $w(z)$ (see equation 5), obtained with window function of width
$\Delta z = 0.6$.
This derivation of $w(z)$ requires of theory of gravity
and the value of $\Omega_0$;
General Relativity has been adopted here, and
$\Omega_0 = 0.3$ is assumed.  The value at
zero redshift is $w_0 = -0.9 \pm 0.1$, consistent with the
cosmological constant models.
}

\end{figure}

\section{Results and Conclusions}

The results presented here follow
those presented by Daly \& Djorgovski (2003, 2004a), where more
details can be found. 
We perform a test of the procedure using a simulated
data set which mimics the anticipated SN measurements from the SNAP/JDEM
satellite (see http://snap.lbl.gov), with a known assumed cosmology,
namely the standard Friedmann-Lemaitre model with $\Omega_0 = 0.3$
and $\Lambda_0 = 0.7$ (see Daly \& Djorgovski 2003, 2004a 
for more details on this simulated
data set).  The results for the dark energy parameters as functions
of redshift are shown in Figure 1.  We see that our method can recover
robustly the assumed parameters, at least out to $z \approx 0.9$.
Reassured by this test, we 
turn to the analysis of actual data.

The
data used here includes 20 radio galaxies (RG)
compiled by Guerra, Daly, \& Wan (2000)
and
the ``gold'' supernova (SN) sample
compiled by Riess et al. (2004).  We note that in the redshift
interval where the two sets of coordinate distances (RG and SN)
overlap, the agreement is excellent (see Figure 2), 
suggesting that neither one
is affected by some significant bias, and allowing us to combine
them for this study.

Measurements of luminosity distances and angular size distances
are easily converted to coordinate distances, $y(z)$, and these
are shown in Figure 3.  The dimensionless expansion
rate obtained from these data using equation (1) 
is shown as a function of
redshift in Figure 4.  The result is consistent with the
standard ``concordance model,'' of $\Omega_0=0.3$ and 
$\Lambda = 0.7$.  The dimensionless acceleration rate
obtained from these data and equation (2) is shown in 
Figure 5. 

We see that the universe transitions from acceleration to deceleration
at a redshift of about 0.4 (consistent with determinations by
Daly \& Djorgovski 2003, 2004a,b, Riess et al. 2004, and Alam, Sahni,
\& Starobinsky 2004); and our determination 
only depends upon the assumption
that the universe is homogeneous, isotropic, and spatially flat.

Assuming GR, we solve
for the pressure (see equation 3 and Figure 6), 
energy density (see equation 4 and Figure 7), 
and equation of state (see equation 5 and Figure 8), 
of the dark
energy.  Each is generally consistent with remaining constant to
a redshift of about 0.5 and possibly beyond, but determining their
behavior at higher redshifts is severely limited by the available data.

As more and better data become available, particularly more coordinate
distance determinations for sources with redshifts between 0.4 and 1.5, 
this methodology can be used
to determine the evolution of the dark energy properties and the
observed kinematics of the universe with an increasing precision
and confidence.

\begin{quote}
{\bfseries Acknowledgments.} We thank the organizers of the workshop, 
especially Tod Lauer, for coordinating 
such an outstanding meeting.  We acknowledge the great
work and efforts of many observers who obtained the valuable data
used in these studies. This work
was supported in part by the U. S. National Science Foundation under 
grant AST-0206002 (RAD) and by the Ajax Foundation (SGD).  
\end{quote}

\end{document}